\documentclass{ws-ijmpcs}

\begin{document}

\markboth{Liuti, Courtoy, Goldstein, Gonzalez Hernandez, Rajan}
{Partonic Picture of GTMDs}

\catchline{}{}{}{}{}
%
%%%%%%%%%%%%%%%%%%%%%%%%%%%%%%%%%%%%%%%%%%%%%%%%%%%%%%%%%%%%%%%%%%%%

\title{PARTONIC PICTURE OF GTMDS
%%\footnote{For the title, try not to use more than 3 lines. Typeset the title in 10 pt roman, uppercase and boldface.}
}

\author{SIMONETTA LIUTI, ABHA RAJAN}

\address{Department of Physics, University of Virginia, \\
Charlottesville, VA 22901, USA.\\
and
Laboratori Nazionali di Frascati, INFN, Frascati, Italy. \\
sl4y@virginia.edu, ar5xc@virginia.edu}

\author{AURORE COURTOY}

\address{IFPA-Institut de Physique, Universite de Liege (ULg), \\
4000 Liege (Belgium) \\
and Laboratori Nazionali di Frascati, INFN, Frascati, Italy. \\
Aurore.Courtoy@ulg.ac.be}

\author{GARY R. GOLDSTEIN
%\footnote{Typeset names in 8 pt roman, uppercase. Use the footnote to indicate the present or permanent address of the author.}
}

\address{Department of Physics and Astronomy, Tufts University,\\
Medford, MA 02155,
USA
%\footnote{State completely without abbreviations, the affiliation and mailing address, including country. Typeset in 8 pt italic.}\\
gary.goldstein@tufts.edu}

\author{J. OSVALDO GONZALEZ HERNANDEZ}
\address{INFN, Sezione di Torino, Italy. \\
joghdr@gmail.com}

\maketitle

\begin{history}
\received{Day Month Year}
\revised{Day Month Year}
\end{history}

\begin{abstract}
We argue that due to parity constraints, the helicity combination of the purely momentum space counterparts of the Wigner distributions -- the generalized transverse momentum distributions --
that describes the configuration of an unpolarized quark in a longitudinally polarized nucleon, 
can enter the deeply virtual Compton scattering amplitude only through matrix elements involving a final state interaction.
The relevant matrix elements in turn involve light cone operators projections in the transverse direction, or they appear 
in the deeply virtual Compton scattering amplitude at twist three.
Orbital angular momentum or the spin structure of the nucleon was a major reason for these various distributions and amplitudes to have been introduced. 
We show that twist three contributions to deeply virtual Compton scattering provide observables related to orbital angular momentum. 
\end{abstract}

\ccode{PACS numbers: 11.25.Hf, 123.1K}
\vspace{0.2cm}

%%%%% INTRO
%%%%%
\section {INTRODUCTION}
\begin{figure}
\hspace{-0.5cm}
%\centerline{
\includegraphics[width=6.0cm]{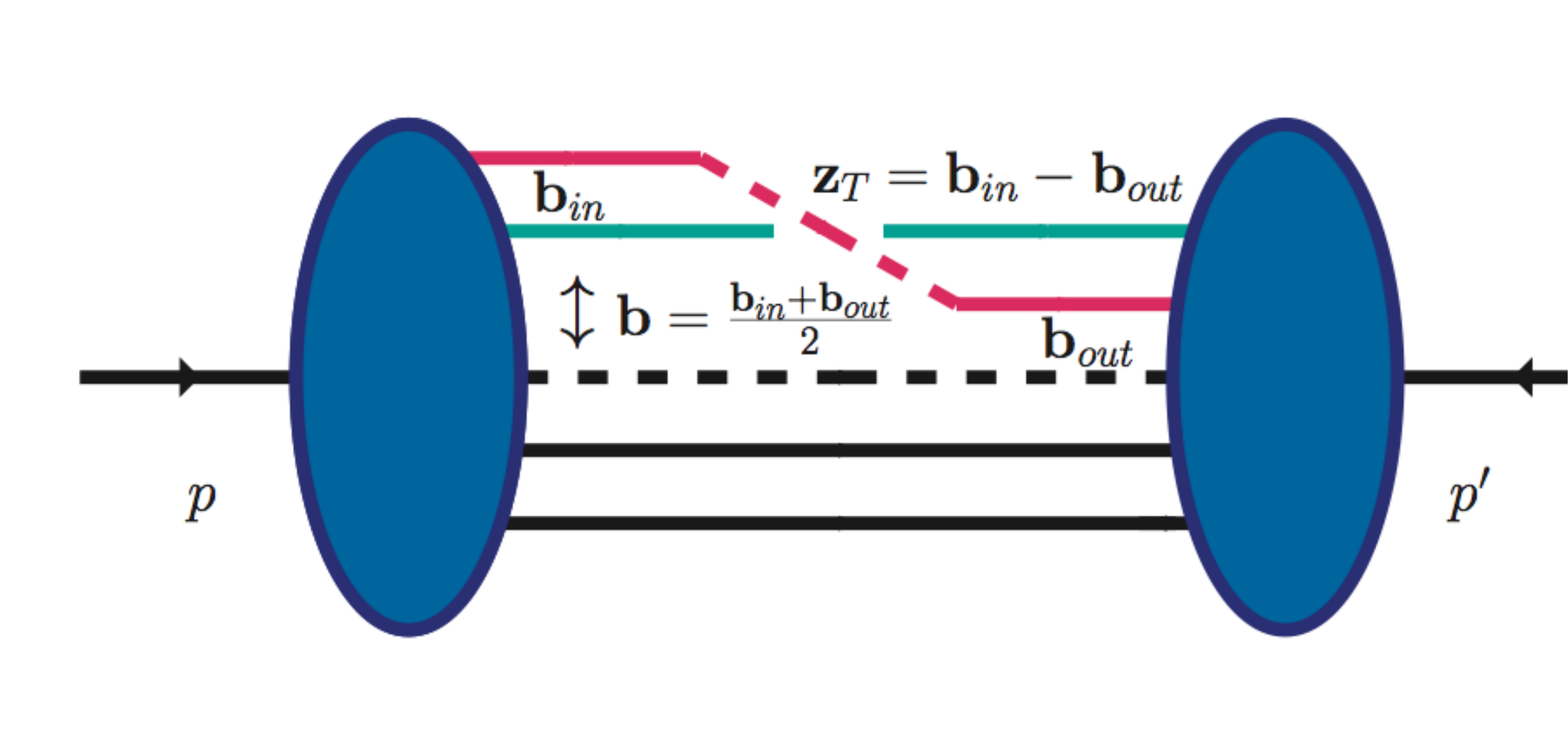}
\hspace{0.5cm}
\includegraphics[width=6.0cm]{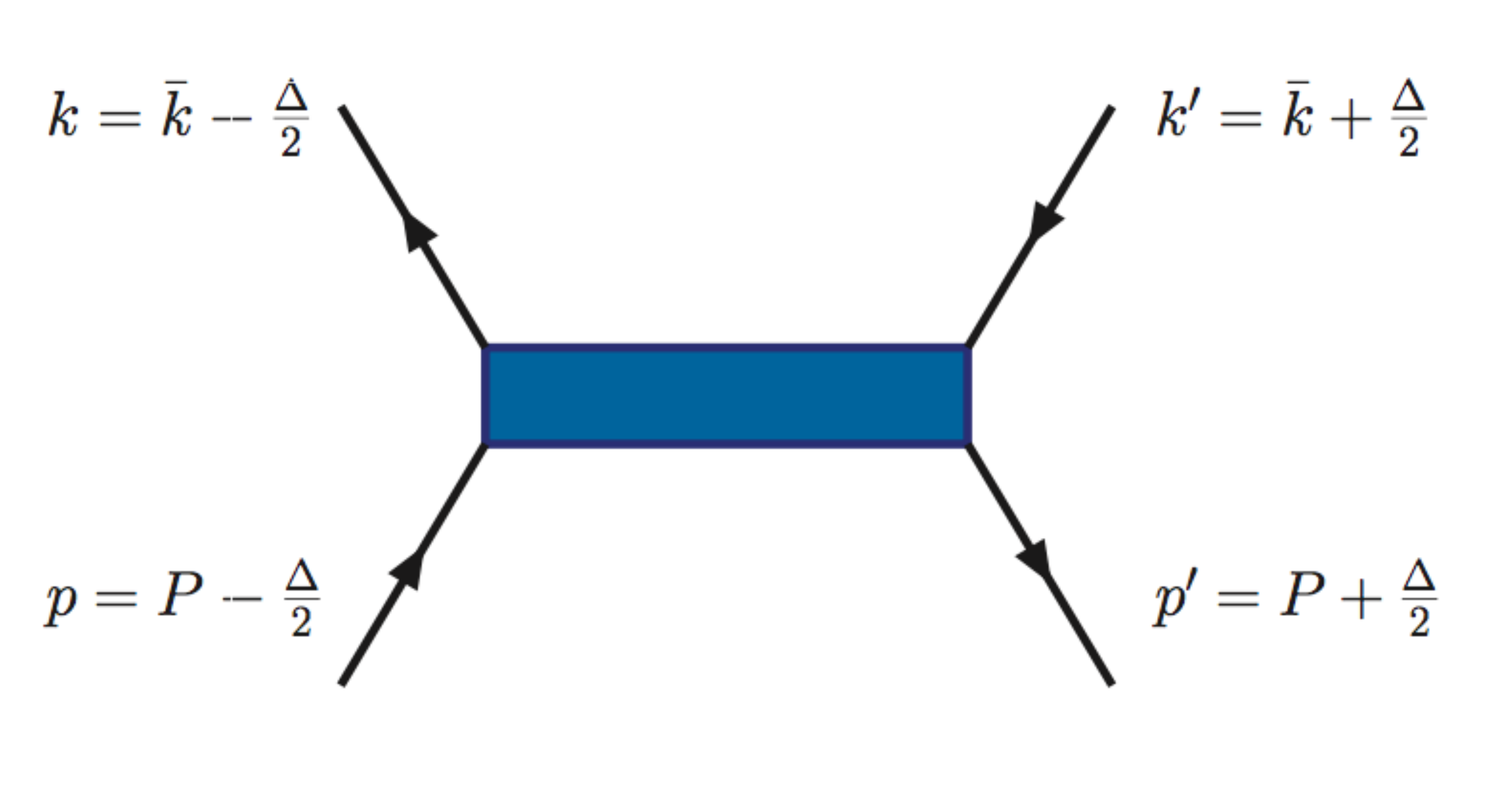}
\caption{{\bf (a)} Left: Correlation function for a GTMD; {\bf (b)} quark-proton scattering in the $u$-channel.}
\label{fig1_f14}
\end{figure}

%
%\noindent {\bf 1.} 
Considerable attention has been devoted to the partons' Transverse Momentum Distributions (TMDs), to the Generalized Parton Distributions (GPDs), and to finding a connection between the two [\refcite{Bur,Metz1,Metz2}]. 
TMDs are distributions of different spin configurations of quarks and gluons within the nucleon whose longitudinal and transverse momenta can be accessed in Semi-Inclusive Deep Inelastic Scattering (SIDIS). 
GPDs are real amplitudes for quarks or gluons being probed in a hard process and then returning to reconstitute a scattered  nucleon. 
%These amplitudes also involve the dynamics of  quarks and gluons confined to the nucleon. 
They are accessed through exclusive electroproduction of vector bosons along with the nucleon. In each case there is a nucleon matrix element of bilinear, non-local quark or gluon field operators. In principle both TMDs and GPDs are different limits of Wigner distributions, {\it i.e.} the  phase space distributions in momenta and impact parameters. The purely momentum space form of those are the Generalized TMDs (GTMDs). 
GTMDs correlate hadronic states with same parton longitudinal momentum, $x$ (assuming zero skewness), different relative transverse distance, ${\bf z}_T$, between the struck parton's initial and final states, and same average transverse distance, ${\bf b}$, of the struck parton with respect to the center of momentum [\refcite{Soper77}] (Figure \ref{fig1_f14}{\bf(a)}).

Understanding the angular momentum or spin structure of the nucleon is a major reason for these various distributions and amplitudes to have been introduced. 
%
%%%
%%% FIGURE 1
%%%
%%%%%%
%%%%%%
%The reason for these different numbers can be traced to the way the kinematics are defined for the GTMDs vs. their reduction to TMDs and GPDs. 
%{\bf To see how the number of amplitudes increases we have to focus on the spinors.} %{\it i.e.} on the dynamics.}
Recently, specific GTMDs and Wigner distributions were studied that are thought to
be related to the more elusive component of the  angular momentum sum rule, which is partonic Orbital Angular Momentum (OAM) [\refcite{Ma,LorPas,Yuan}]. Such theoretical efforts have been developing in parallel with  the realization that the leading twist contribution to the angular momentum sum rule comes from transverse spin [\refcite{Ji_Xiong}], while longitudinal angular momentum, and consequently orbital angular momentum, can be associated with twist three partonic components.
The GTMD that was proposed to describe OAM appears in the parametrization of the vector, $\gamma^+$, component of the unintegrated quark-quark correlator for the proton given in [\refcite{Metz2}] as, \[ \bar{u}(p',\Lambda') \frac{i \sigma^{ij} \bar{k}_T^i \Delta_T^j}{M^2} u(p,\Lambda) F_{14} \]
where $(p, \Lambda), (p',\Lambda')$ are the proton's initial and final momentum and helicity, $k_T$ and $\Delta_T$ are the quarks' average and relative momenta, respectively, and $F_{14}$ is the GTMD defined according to the classification scheme of Ref.[\refcite{Metz2}].  
Although this term is suggestive of OAM, it is inconsistent with several physical properties, namely: 

\noindent  {\it  i)} it drops out of the formulation of both GPDs and TMDs, so that it cannot be measured;

\noindent {\it ii)} it is parity-odd;

\noindent {\it iii)} it is non zero only for imaginary values of the quark-proton helicity amplitudes.      

In order to develop a more concrete understanding of OAM that could lead to the definition of specific observables, it is important to 
develop a physical sense for the newly proposed partonic configurations and of their connection with the quark-proton helicity or transverse spin amplitudes. 
In this paper, after examining the constraints on the recently studied GTMDs from the invariance under parity transformations, 
we perform a thorough analysis of their helicity/ transverse spin structure, and we suggest that such terms can be non zero only at twist  three. 
%We subsequently verify our calculations using models. 
Our findings corroborate the sum rule originally proposed in Ref.[\refcite{Polyakov}] (see also Ref.[\refcite{Hatta}]). They bear important consequences for the experimental access to orbital angular momentum.

%\vspace{0.3cm}
%%%%%% 
%\noindent {\bf 2.} 
\section{Leading Twist GTMDs}
%The Wigner distributions for deeply virtual electroproduction were introduced by X. Ji and were advocated as an approach to an holographic view of the partons within the nucleon. Different combinations of spin amplitudes could reveal the sharing of spin and orbital angular momenta among the quarks and gluons. Extending this  approach through amplitudes or distributions in the parton momenta and spin orientations defines the GTMDs. 
In Ref.[\refcite{Metz2}] it was found that with parity invariance, time reversal invariance, and Hermiticity there are 16 independent complex GTMDs for the quark-nucleon system, corresponding to 16 helicity amplitudes for quark-nucleon elastic scattering. 
%This number of independent GTMDs is to be compared with 8 real leading twist GPDs or TMDs.  Since the quarks and nucleons each have two spin states, there are 16 combinations for $q^\prime + N \rightarrow q + N^\prime$. Parity conservation reduces that number by half for either TMDs or GPDs. So the number 16 for independent GTMDs is somewhat surprising, The GTMDs certainly satisfy Parity conservation. How is that implemented while not reducing the number from 16 to 8? 
However, we know that for elastic 2-body scattering of two spin $1/2$ particles there will only be 8 independent amplitudes. This follows from implementing Parity transformations on the helicity amplitudes in the 2-body Center of Mass (CM)  frame [\refcite{JacobWick}] where all the incoming and outgoing particles are  confined to a plane. In this plane the Parity transformation flips all momenta but it does not change the relation among the momentum components. In any other frame there will still be 8 independent amplitudes, although they may be in linear combinations with kinematic factors that {\em appear} to yield 16  . 
The counting of helicity amplitudes in polarization dependent high energy scattering processes was addressed {\it e.g.} Ref.[\refcite{FGM}]. In order to explain this point, and to investigate its consequences on the  off-forward matrix elements of QCD correlators, we first start by reviewing the helicity structure of the GTMDs from Ref.[\refcite{Metz2}].

To describe quark-proton scattering as a $u$-channel  two-body scattering process (Fig.\ref{fig1_f14}{\bf (b)})
\[ q^\prime(k^\prime) + N(p) \rightarrow q(k) + N^\prime(p^\prime),  \] 
we choose a Light-Cone (LC) frame, where the {\em average} and {\em relative} 4-momenta are $P=(p+p')/2$, $\bar{k}= (k+k')/2$, $\Delta = p'-p= k'-k$, respectively.  We take the skewness variable, $\xi=0$ since this will not enter our discussion.

The unintegrated matrix elements defining the GTMDs and the quark-proton helicity amplitudes can be connected using the following definition, 
\begin{eqnarray}
A_{\Lambda' \lambda', \Lambda \lambda} = \int \frac{d z^- \, d^2{\bf z}_T}{(2 \pi)^3} e^{ixP^+ z^- - i{\bf k}_T\cdot {\bf z}_T} \left. \langle p', \Lambda' \mid {\cal O}_{\lambda' \lambda}(z) \mid p, \Lambda \rangle \right|_{z^+=0}, \nonumber \\
\end{eqnarray} 
%
%%% p.2
%Notice that particles off-shellness is a matter of concern only for GTMDs and not for TMDs and GPDs: 
%for TMDs  there is no ${\bf \Delta}$ -  the initial and final quark and hadron have the same momenta, therefore all four-particles belong to the same plane at leading twist; 
%for GPDs the off-shellness, $k^2$,  is effectively integrated over by transforming from the $k_T$ variable keeping the ``external" variables fixed. 
where in the chiral even sector, 
\begin{eqnarray}
\label{tw2_operator}
{\cal O}_{\pm \pm}(z) & = & \bar{\psi}\left(-\frac{z}{2}\right) \gamma^+(1 \pm \gamma_5)  \psi\left(\frac{z}{2}\right). 
%{\cal O}_{- -}(z) & = & \bar{\psi}(-\frac{z}{2}) \gamma^+(1-\gamma_5)  \psi(\frac{z}{2}), 
\end{eqnarray}
%and in the chiral odd sector,
%\begin{eqnarray}
%{\cal O}_{-+}(z) & = & -i \bar{\psi}(-\frac{z}{2}) (\sigma^{+1} - i \sigma^{+2})  \psi(\frac{z}{2}) \\
%{\cal O}_{+ -}(z) & = & i \bar{\psi}(-\frac{z}{2}) (\sigma^{+1} + i \sigma^{+2}) \psi(\frac{z}{2}), 
%\end{eqnarray}
%so that, 
%\begin{eqnarray}
%A_{\Lambda \pm, \Lambda' \pm} & \rightarrow  & \langle  p', \Lambda' \mid \bar{\psi}(-\frac{z}{2}) \gamma^+(1\pm \gamma_5)  \psi(\frac{z}{2}) \mid p, \lambda \rangle 
%A_{\Lambda -, \Lambda' -} & \rightarrow  & \langle  p', \Lambda' \mid \bar{\psi}(-\frac{z}{2}) \gamma^+(1-\gamma_5)  \psi(\frac{z}{2}) \mid p, \lambda \rangle 
%A_{\Lambda -, \Lambda' +} & \rightarrow  & \langle  P', \Lambda' \mid \bar{\psi}(-\frac{z}{2}) (\gamma^1+i \gamma^2)  \psi(\frac{z}{2}) \mid P, \lambda \rangle \\
%A_{\Lambda +, \Lambda' -} & \rightarrow  & \langle  P', \Lambda' \mid \bar{\psi}(-\frac{z}{2}) (\gamma^1- i \gamma^2)  \psi(\frac{z}{2}) \mid P, \lambda \rangle.
%\end{eqnarray}
%%
%%
Using the notation of Ref.[\refcite{Metz2}], one finds
the new contributions, $F_{14}$, and $G_{11}$, that appear as linear combinations
\begin{subequations}
\begin{eqnarray}
\label{F14}
 i  \frac{\bar{k}_1\Delta_2 - \bar{k}_2 \Delta_1}{M^2} F_{14} & = & A_{++,++} + A_{+-,+-} - A_{-+,-+} - A_{--,--} \\
 \label{G11}
 i  \frac{\bar{k}_1\Delta_2 - \bar{k}_2 \Delta_1}{M^2}  G_{11} & = & A_{++,++} -A_{+-,+-} + A_{-+,-+} - A_{--,--}   
\end{eqnarray}
\end{subequations}
$F_{14}$ describes an unpolarized quark in a longitudinally polarized proton, while $G_{11}$ describes a longitudinally polarized quark in an unpolarized proton.

%For a two-body scattering process, these distributions clearly violate parity. 

Parity, however,  imposes limits on the possible polarization asymmetries that can be observed in two body scattering:    
because of 4-momentum conservation and on-shell conditions, $k^2=m^2$, $p^2=M^2$, there are eight variables. Four of those  describe the energy and 3-momentum of the CM relative to a fixed coordinate system, while the remaining four give the energy  and the 3-vector orientation and magnitude of the scattering plane in the CM.
In the CM frame or, equivalently in  the ``lab" frame with the $p$ direction chosen as the $z$-direction, 
%we can now ask whether there can be a measurable polarization of the struck quark in the longitudinal direction. 
the net longitudinal polarization 
%For our spin 1/2 target (or beam) that is the question: ``can there be a non-zero expectation value of 
 is defined by $({\bf \sigma} \cdot {\bf k})$ which is clearly a parity violating term (pseudoscalar) under space inversion (${\bf k} \rightarrow - {\bf k}$). 
 This implies that a measurement of  {\em single longitudinal polarization asymmetries} would violate parity conservation in an ordinary two body scattering process corresponding to tree level, twist two amplitudes. We can therefore anticipate that similarly to the TMDs $g^\perp, f_L^\perp, \ldots $ in SIDIS, single longitudinal polarization asymmetries are higher twist objects. 
%However it seems to appear at leading twist in the GPCFs and GTMDs in MMS. 
On the other hand, notice that polarization along the normal to the scattering plane $[{\bf \sigma} \cdot ({\bf k} \times {\bf p}^\prime)]$ is parity conserving under spatial 
inversion, thus giving rise to SSAs at leading twist [\refcite{JacRobRos}]. 
%. 
 
Because of the parity constraints [\refcite{Metz2}]  $F_{14}$ can therefore be non zero only if its corresponding helicity amplitudes combination is imaginary. Hence it cannot have a straightforward partonic interpretation. Integrating over $k_T$ gives zero for $F_{14}$ meaning that this term decouples from partonic angular momentum sum rules (details of the calculation will be given in Ref.[\refcite{inprogress}]). 
We conclude that this term should not be included in the leading twist parametrization. A similar argument is valid for the axial vector component. 

%\vspace{0.3cm}
\section{Twist three}
%%%%% 
%%%%% SECTION 3
%%%%%
%\noindent {\bf 3.} %HYPOTHESIS OF TW 3
%How is it that both $F_{14}$ and $G_{11}$ seem to be different from zero when calculated within models (specifically the scalar diquark model in Ref.\refcite{Metz2}, and the constituent quark model in \refcite{LorPas})?  
%
%%%% OFF-SHELL STATEMENT
%Next we demonstrate that while the number of  amplitudes will not be modified by kinematics, or by the fact that some of the particles can be 
%off-mass shell, more amplitudes can instead be generated by  relaxing  the on-shell constraints on the spin vectors.
%The new amplitudes which are present because of the transverse components connected to off-shellness, correspond to contributions off the LC direction,     
%or of higher twist. This is important for the physical interpretation of the quark Wigner distributions implying, in particular, that distributions of longitudinally polarized quarks in an unpolarized nucleon, $\rho_{LU}$ and $\rho_{UL}$,  require the presence of a final state interaction to be meaningful.  
%
%
%Viewing  the GTMDs as originating from two body scattering in the $u$-channel with parton  off-shellness one has that  the transverse vectors $\vec{k}_T$ and $\vec{\Delta}_T$ can now vary independently as if there were another particle with its own separate momentum. 
In the presence of final state interactions parity relations apply differently. 
%as if there were a third particle. 
%
%\[ 
%A_{++,++} + A_{+-,+-} - A_{--,--} - A_{-+,-+}. \]
%This contribution derives from the spinor components (the ``bad" components) that are possible because of off-shellness. 
%%%%%
%%%%% DETAILED DESCRIPTION OF TW 3
%%%%%
The chiral-even twist three components were also parametrized in Ref.[\refcite{Metz2}],
%%% SL 5/7
\begin{eqnarray}
\label{tw3_metz}
W^{\gamma^i}_{\Lambda' \Lambda} & = & \frac{1}{2 P^+} \overline{U}(p',\Lambda')  \left[% 
\frac{\bar{k}_T^i}{M}  F_{21}  + 
\frac{\Delta_T^i}{M}  F_{22}  
 +   \frac{ i\sigma^{ji}\bar{k}_T^j}{M}  F_{27}  +  \frac{ i\sigma^{ji}\Delta_T^j}{M}  F_{28}  + \frac{M i\sigma^{i+}}{P^+}  F_{23} \right. \nonumber \\
 & + &  \left.   \frac{\bar{k}_T^i}{M} \frac{ i\sigma^{k+}\bar{k}_T^k}{P^+}  F_{24} + \frac{\Delta_T^i}{M} \frac{ i\sigma^{k+}\bar{k}_T^k}{P^+} F_{25}
+   \frac{\Delta_T^i}{M} \frac{ i\sigma^{k+}\Delta_T^k}{P^+}  F_{26}  \right]    U(p,\Lambda) 
 \end{eqnarray} 
where $i=1,2$. A similar decomposition applies to the axial counterpart.
The twist three correlators involve composite systems of a transverse gluon and quark for which helicity and chirality are opposite. Since the gluon carries helicity but no chirality, by imposing angular momentum conservation one obtains the opposite chirality [\refcite{KogSop,Jaffe}]. 
This leads to the following expression for the helicity amplitude,
\begin{eqnarray}
\label{helamp_tw3}
A^{tw 3}_{\Lambda' \lambda', \Lambda \lambda} & = \displaystyle\int \frac{d z^- \, d^2{\bf z}_T}{(2 \pi)^3} e^{ixP^+ z^- - i{\bf k}_T\cdot {\bf z}_T} \left. \langle p', \Lambda' \mid {\cal O}_{-\lambda' \lambda}(z) \mid p, \Lambda \rangle \right|_{z^+=0}, 
\end{eqnarray} 
where  ${\cal O}_{-\lambda' \lambda}(z)$ is the twist three quark field operator.
\begin{subequations}
\label{tw3_operator2}
\begin{eqnarray}
{\cal O}^q_{+ -}(z) & = & \phi^\dagger_{+}\left(-\frac{z}{2}\right) \chi_{+}\left(\frac{z}{2}\right)  - \chi^\dagger_{-}\left(-\frac{z}{2}\right) \phi_{-}\left(\frac{z}{2}\right)   \\
{\cal O}^q_{- +}(z) & = & \chi^\dagger_{+}\left(-\frac{z}{2}\right) \phi_{+}\left(\frac{z}{2}\right)  - \phi^\dagger_{-}\left(-\frac{z}{2}\right) \chi_{-}\left(\frac{z}{2}\right)  
\end{eqnarray}
\end{subequations}
(similar expressions are obtained for the axial vector case).
%
%Note that for the non flip quark-proton helicity amplitudes at twist three one finds that the chiral even structures correspond to what would be chiral odd at twist two,
%\begin{eqnarray}
%A^{ tw 3}_{\Lambda' \pm, \Lambda \pm}  \rightarrow A^{tw 2}_{\Lambda' \pm, \Lambda \mp}.
%A_{\Lambda -, \Lambda' -} & \rightarrow  & \langle  p', \Lambda' \mid \bar{\psi}(-\frac{z}{2}) \gamma^+(1-\gamma_5)  \psi(\frac{z}{2}) \mid p, \lambda \rangle 
%\end{eqnarray}
 
 The helicity amplitudes combinations describing the unpolarized quark in a longitudinally polarized proton, 
yield upon integration over $k_T$,  the twist three GPD $\widetilde{E}_{2T}$. This was identified with OAM in Ref.[\refcite{Polyakov}], 
%BelMuel,BelRad},
\begin{eqnarray}
\widetilde{E}_{2T} & = & -2 \int d^2 {\bf k}_T  \left[ \left(\frac{{\bf k_T} \cdot {\bf \Delta}}{\Delta_T^2} \right) F_{27} +  F_{28}\right]  \rightarrow G_2 
\label{G2} 
\end{eqnarray}
where $G_2$ is obtained from [\refcite{Polyakov,BelMuel,BelRad}],
\begin{equation}
F_\perp^\mu=G_1\frac{\Delta_\perp^\mu}{2M}+G_2\gamma_\perp^\mu+G_3\Delta_\perp^\mu\gamma^+ +G_4i\epsilon_{\mu\nu}\Delta_\perp^\mu\gamma^+\gamma^5
\end{equation}

{\em The twist 3 expressions should therefore replace the twist 2 combination corresponding to  $F_{14}$, Eq.(\ref{F14}) in the interpretation of OAM}. 

\noindent We know, in fact, that $F_{14}$ decouples from physically measurable quantities. 
The twist three GPDs enter the observables as Compton Form Factors in combination with twist two as [\refcite{BelMuel}],
\begin{eqnarray}
{\cal F }^{eff} = -2\xi \left( \frac{1}{1+\xi} {\cal F} + {\cal F}^3_+ - {\cal F}^3_- \right)   
\end{eqnarray} 
with ${\cal F} = {\cal H}, {\cal E},...$. In turn these Compton Form Factors (CFFs) enter the DVCS  cross section terms as magnitude squares 
and the Bethe Heitler interference term as EM form factors times CFFs.
The GPD of interest is $G_2$, so we could already find its signature in upcoming data analyses [\refcite{pisano_privcomm}]. 

\section{Angular momentum sum rules}
%\vspace{0.3cm}
%%%%
%\noindent {\bf 4.} 
We now show how the results obtained in the previous Section determine the contribution of OAM to the proton's angular momentum sum rule derived in Refs.[\refcite{JM_SR,Ji_SR}]. While the derivation of the sum rule was carried out along similar lines in both Refs.[\refcite{JM_SR}] and [\refcite{Ji_SR}], the two approaches essentially differ in that in Ref.[\refcite{JM_SR} ](JM) one has,
\begin{equation}
\frac{1}{2} =  \frac{1}{2} \Delta \Sigma + {\cal L}_q + \Delta G + {\cal L}_g,
\end{equation} 
where ${\cal L}_{q(g)} \rightarrow {\bf r} \times i{\bf \partial}$, {\it i.e.}  corresponds to  canonical OAM, while in Ref.[\refcite{Ji_SR}] (Ji),
\begin{equation}
\label{Ji_SR:eq}
\frac{1}{2} = J_q + J_g =  \frac{1}{2} \Delta \Sigma + L_q + J_g,
\end{equation}
where $L_q \rightarrow  {\bf r} \times i{\bf D}$ includes dynamics through the covariant derivative. Furthermore, $J_g$, the gluons total angular momentum contribution to Eq.(\ref{Ji_SR:eq}) cannot be split into its separate intrinsic and orbital components, in order to satisfy gauge invariance.  In Ref.[\refcite{Ji_SR}]  the quarks and gluons angular momentum components were identified with observables obtained from Deeply Virtual Compton Scattering (DVCS) type experiments. Both $J_{q(g)}$ and $L_q$ can therefore be measured owing to the well known relation, 
\begin{eqnarray}
\int_{-1}^1 dx \, x(H_{q(g)}(x,0,0) &+&E_{q(g)}(x,0,0)) = J_{q_(g)} \rightarrow \nonumber \\
L_q = \int_{-1}^1 dx \, &x& (H_{q}(x,0,0)+E_{q}(x,0,0)) - \int_{-1}^1 dx \, \widetilde{H}(x,0,0)
\end{eqnarray}

What is crucial here is that in a subsequent development Polyakov {\it et al.} [\refcite{Polyakov}] derived a sum rule for the twist three vector components,
\begin{eqnarray}
\label{polyakov:eq}
\int dx \, x \, G_2^q(x,0,0)  =  \frac{1}{2} \left[ - \int dx  x  (H^q(x,0,0) + E^q(x,0,0)) + \int dx \tilde{H}^q(x,0,0) \right] 
% \frac{1}{2}  \int dx \, x (H(x,0,0) + E(x,0,0))  & = & J^q \\
% \int dx \tilde{H}^q(x,0,0)  & = & \Sigma^q 
\end{eqnarray}
from which they deduced that the second moment of $G_2$ represents the quarks' OAM.
By connecting this result with the twist three helicity amplitudes derived in Section {\bf 3.} we therefore conclude that the distribution of an unpolarized quark in a  longitudinally polarized nucleon does indeed measure OAM as the intuitive arguments in Refs.[\refcite{LorPas,Yuan}] suggested. However, this must be identified with a twist three contribution.
Our finding is in line with the recent observation that the same twist three contribution, whose appearance had already been noticed in Ref.[\refcite{Polyakov}], is fundamental for solving the issue of defining the quarks and gluons angular momentum  decomposition within QCD [\refcite{Ji_Xiong,Hatta}].

In order to connect the partonic interpretation of OAM  in Eq.(\ref{polyakov:eq}) which relates directly to the Ji sum rule, with the JM decomposition, one needs to examine in detail the nature of the twist three contributions. We find that OAM according to the JM decomposition also appears at twist 3 and cannot be therefore identified with $F_14$. The results of this analysis will be presented in another publication [\refcite{inprogress}].

\section{Conclusions}
%\vspace{0.3cm}
%%%%%%% 
%\noindent {\bf 5.}  
In conclusion, 
as a result of two-body kinematics and four-momentum conservation, parity conservation in the Center of Momentum frame (CoM) for quark-nucleon scattering limits the number of independent helicity amplitudes to eight - four chiral even and four chiral odd. If each GTMD is considered at fixed values of its arguments (with $\xi=0 $), then there will be a Lorentz transformation that depends on those fixed values to bring the kinematics into a single CM plane. In that plane there will be eight (complex) independent amplitudes. Hence the apparent sixteen GTMDs at leading twist reduce to eight TMDs in one limit and eight GPDs in another. In neither of those limits do $F_{14}$ or $G_{11}$ survive. They are thus not observable. Nevertheless in various models these GTMDs are non-zero. It appears that these non-zero results are coming about from the kinematics or from effective higher twist components arising from quarks' confinement. There can not be leading twist CM amplitudes with the Dirac and $k_T, \Delta_T$ kinematic factors in the CM - the two transverse momenta become planar and they have the wrong parity signatures. 

Starting from this observation we proposed a QCD approach where: 1) single longitudinal polarizations observables can be derived; 2) they involve twist three distributions. Our approach is complementary to the one in Ref.[\refcite{Burkardt_torque}] that was derived using TMD factorization.

Our most important result is perhaps in dispelling the notion that what is believed to be the orbital angular momentum component of the nucleon spin sum rule cannot be observed directly in hard scattering experiments. Both the JM and Ji decompositions correspond to twist three contributions, and their validity can be tested by measuring twist three GPDs. Some of these observables might already be attainable from recent accurate Jefferson Lab measurements [\refcite{pisano_privcomm}].   

This work was supported by the U.S. Department of Energy grant DE-FG02-01ER4120.

\end{document}